\documentclass[twocolumn,prl,aps,showpacs,superscriptaddress]{revtex4}
\usepackage{graphicx}
\usepackage{amssymb}
\usepackage{textcomp}
\def\be{\begin{equation}}
\def\ee{\end{equation}}
\def\bea{\begin{eqnarray}}
\def\eea{\end{eqnarray}}
\def\bdm{\begin{displaymath}}
\def\edm{\end{displaymath}}
\def\ba{\begin{array}}
\def\ea{\end{array}}

\begin{document}

\title{Magnetic Breakdown in Twisted Bilayer Graphene}

\author{Chi-Ken~Lu}
\author{H. A. Fertig}
\affiliation{Department of Physics, Indiana University, Bloomington, Indiana 47405, USA}
\date{\today }

\begin{abstract}

We consider magnetic breakdown in twisted bilayer graphene where electrons may hop between semiclassical $k$-space trajectories in different layers. These trajectories within a doubled Brillouin zone constitute a network in which an $S$-matrix at each saddle point is used to model tunneling between different layers. Matching of the semiclassical wavefunctions throughout the network determines the energy spectrum. Semiclassical orbits with energies well below that of the saddle points are Landau levels of the Dirac points in each layer.  These continuously evolve into {\it both} electron-like and hole-like levels above the saddle point energy. Possible experimental signatures are discussed.


\end{abstract}


\maketitle

\emph{Introduction} -- The fundamental description of electron dynamics in a crystal
and a uniform magnetic field involves orbital motion in a plane perpendicular
to the field, along contours of constant energy \cite{AM,LandauSP2} as a function
of crystal momentum {\bf k}.
This behavior can be significantly modified when tunneling from
one trajectory to another becomes important, a phenomenon
known as magnetic breakdown (MB) \cite{MB1,Pippard,Chambers}.  MB is important when the closest
approach between $k$-space trajectories is on the order of the
inverse of magnetic length,
$\ell_B\approx 100/\sqrt{B[T]}$
nm, where $B(T)$ is the magnetic field in Tesla.
MB sometimes leads to the
formation of open orbits, with dramatic transport signatures \cite{AM}.


MB effects in bulk metals can be challenging to observe because
saddle points in a band structure, where MB initially sets in as
the electron energy changes \cite{MB1}, are often quite far from
the Fermi energy. Recently, excellent candidates to observe MB
phenomena have become available in the form of twisted graphene
bilayers \cite{CBerger,Andrei} and graphene deposited on boron
nitride substrates \cite{hBN1,hBN2,hBN3}. These two-dimensional
systems can support large unit cells in real space (``Moir\'e
patterns''), and correspondingly small Brillouin zones, for which
critical points in the energy dispersion can be at relatively low
energy \cite{MB2,Santos1}. Such large unit cells have allowed the
recent observation of the self-similar Hofstadter
spectrum \cite{Hofstadter,Hofstadter_exp1,Hofstadter_exp2}, and may
in principle nucleate unusual many-body
states \cite{levitov,Gonzalez,Heinz} for Fermi energies near that
of a saddle point. These interesting behaviors are among the
reasons that twisted bilayers have attracted so much
attention \cite{Mele,Shallcross,nonabelian,deGail1,Choi,Bistritzer2,
Moon,QHE_ex1,Brihuega,Ohta,HeLin,Raman1,Raman2}.

A twisted bilayer graphene system is characterized by a rotation
angle $\theta$ of the layers' principal axes relative to an
AA-stacked bilayer. In momentum space, this relative rotation
separates the Dirac points associated with each layer by a
distance $k_{\theta}$. At low magnetic field and for energies just above
those of the Dirac points, momentum space trajectories are
circular and surround one or the other Dirac point. Allowed areas
enclosed by these trajectories are quantized in units of
$1/\ell_B^2$, are electron-like (increase in energy with field), and yield
a spectrum essentially the same
as for uncoupled layers. At higher energy these
trajectories approach one another, and interlayer tunneling
becomes qualitatively important. To understand how the spectrum
evolves it is crucial to recognize that the coupling results in
{\it three} distinguishable, degenerate saddle points. In the
presence of the field, as the energy is raised above that of the
saddle points the semiclassical orbits break apart and reconnect.
The new orbits are topologically distinct from the lower energy
ones in that they enclose \emph{neither} of the Dirac points.
Instead they surround a local maximum and na{\"i}vely should be hole-like,
{\it i.e.} decrease in energy with field \cite{Moon}.
This suggests a very large accumulation of levels at the saddle point energy
at high field. Below we demonstrate
by a careful treatment of the magnetic translational symmetries
that such singular behavior is avoided.
The spectrum
necessarily contains {\it both} hole-like and electron-like
orbits above the saddle point (see Fig. 2), with the latter sweeping
the levels to high energy at large field.  We expect this mechanism
to be generic to band structures in which there is a sharp
transition energy between hole-like and electron-like
semiclassical orbits.

\begin{figure}
\input{epsf}
\includegraphics[width=0.4\textwidth]{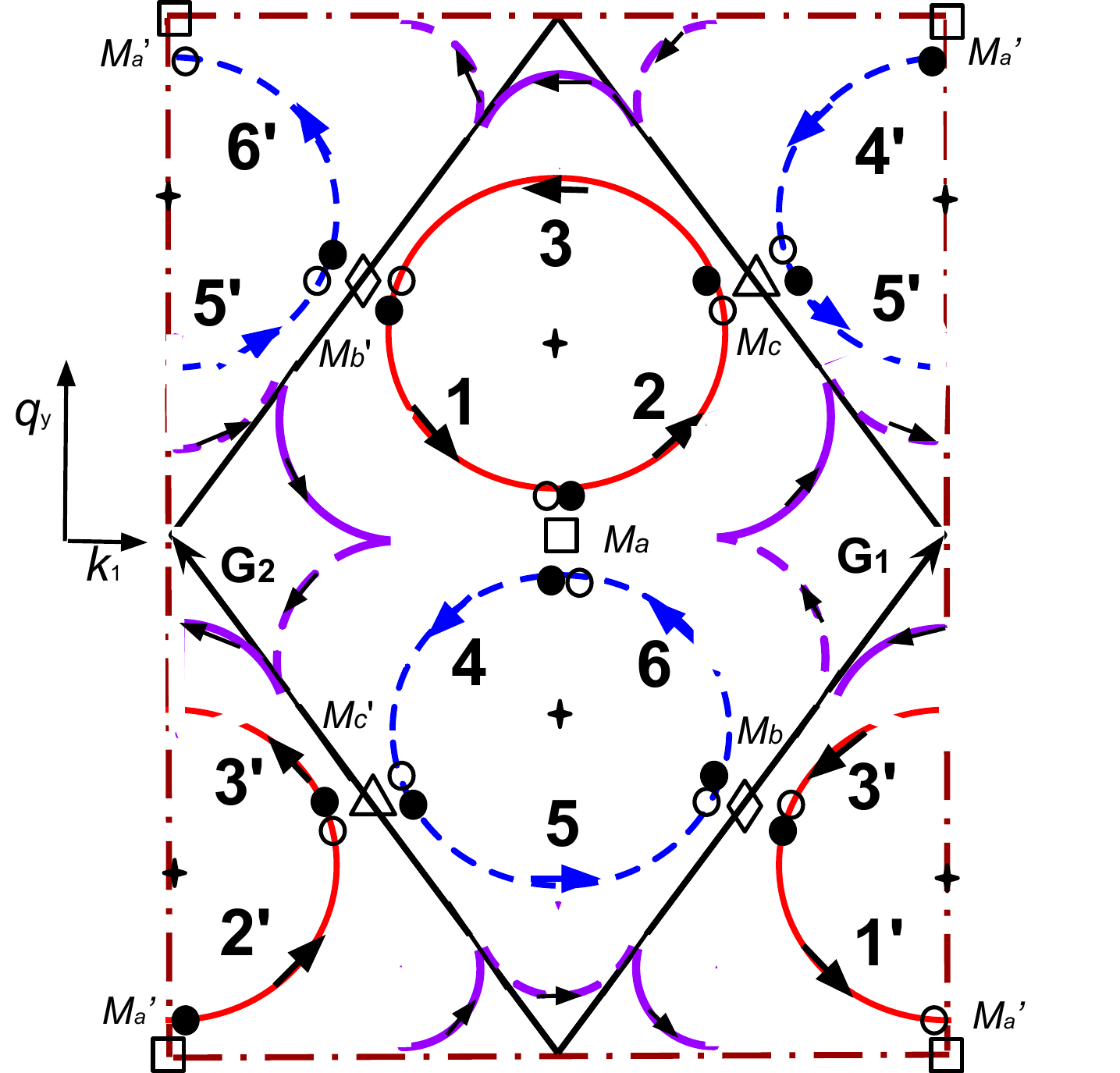}
\caption{(Color online) Semiclassical orbits in the doubled BZ.  Solid (red) trajectories are in
top layer, dashed (blue) are on bottom.  Circular orbits correspond to energies below saddle point,
star-like (purple) orbits are above.  Saddle points are labeled by $M_a$, $M_b$, $M_c$. Symbols of cross represent the Dirac points.
}\label{2BZ}
\end{figure}

The model we adopt for the system, first introduced in Ref. \
\onlinecite{Santos1}, contains one Dirac Hamiltonian for each
layer and three interlayer hopping terms, two of which contain
scattering momenta $\vec G_1$ and $\vec G_2$.  These become
reciprocal lattice vectors for the system (see Fig.\ \ref{2BZ}),
and define the Brillouin zone (BZ) for the zero field energy
spectrum.  In the presence of a magnetic field, energy eigenstates
can simultaneously be eigenstates of two magnetic translation
operators, but, as we show below, the resulting states can be
represented as $k$-space trajectories only if one includes a
minimum of two BZ's in the representation.  This turns out to be a
crucial element in understanding the energy spectrum above the
saddle point: one finds two ``star-like'' semiclassical orbits,
illustrated in Fig.\ \ref{2BZ}.  A very unusual property of these
orbits is that they involve periodic oscillations of the electrons
between layers, and their quantization conditions leads to the
interpenetrating electron- and hole-like levels. We discuss
possible experimental consequences of these properties below.

\vfill\break

{\it Hamiltonian and Saddle Point Dispersions} --
Our starting point is the zero-field Hamiltonian \cite{Santos1,Bistritzer1}

\be
 H=\left(\begin{array}{cccc}
    H_T & w\sum_{i=0,1,2}V_i   \\
    w\sum_{i=0,1,2}V_i^{\dag} & H_B\end{array}\right)\:,\label{BasicModel}
\ee
in which $H_{T,B}=v_F\left[\hat\sigma_x {p}_1+\hat\sigma_y({p}_2\mp\frac{k_{\theta}}{2})\right]$
are the Dirac Hamiltonians for uncoupled top/bottom layers, with Dirac points located at
$\vec k=(0,\pm \frac{k_{\theta}}{2})$, ${p}_{1,2}$ are components
of the momentum operator, and
the Pauli matrices $\hat{\sigma}_{x,y,z}$ act on the sublattice index.
The coupling terms, $\hat{V}_0=\hat{t}_0$,
$\hat{V}_1=\hat{t}_1e^{i\vec{G}_1\cdot\vec{r}}$, and
$\hat{V}_1=\hat{t}_2e^{i\vec{G}_2\cdot\vec{r}}$,
are the largest interlayer hopping terms expected in a continuum model \cite{Santos1,Santos2}.
These introduce discrete translational symmetry characterized by reciprocal lattice vectors
$\vec G_{1,2}=k_{\theta}(\pm\frac{\sqrt{3}}{2},\frac{3}{2}) \equiv (\pm G_x,G_y)$.
The hopping matrices are then specified
by $\hat{t}_0=\hat{\mathbb{I}}_2+\hat\sigma_x$,
$\hat{t}_1=\bar z e^{i\frac{\pi}{3}\hat\sigma_z}\hat{t}_0e^{-i\frac{\pi}{3}\hat\sigma_z}$, and $\hat{t}_1=\hat{t}_2^*$. Here $\hat{\mathbb I}_2$ is the two-dimensional unit matrix, $z=e^{i3\pi/2}$, and $\bar z$ is its complex conjugate.


To understand the behavior of this system,
we treat the interlayer hopping as a weak periodic perturbation.
This has important qualitative effects for nearly degenerate states
in the top and bottom layers that are coupled by the perturbation.
For example (see
Fig.\ \ref{2BZ}), a degeneracy between the top and bottom Dirac bands
in the neighborhood of $M_a$ is split by the interlayer term $V_0$.
Setting $v_F$ to unity, we find that the two states at $k=0$ with energy
$E=\frac{k_{\theta}}{2}$ in the absence of $V_0$ splits into states of energies ${\sqrt{k_{\theta}^2+4w^2}}/{2}\pm w$. Near $M_a$, one
may treat terms involving (small) momenta ${\bf k}=(k_1,k_2)$ perturbatively, to
obtain a two-band effective Hamiltonian \cite{details}
\be
    H_{sp}=({k_{\theta}}/{2}+{k_1^2}/k_{\theta})\hat{\mathbb I}_2+
    \left[\ \alpha(2k_1+k_{\theta})\hat\sigma_z-k_2\hat\sigma_x\ \right]
    \:,\label{2band}
\ee
where $\alpha=w/k_{\theta}$ is small. The eigenstates of $H_{sp}$ include a parabolic band at higher energy, and
a lower band with a saddle point (SP) for which the dispersion is
$\mathcal{E}_{sp}(k_1,k_2)={k_{\theta}}/{2}+{k_1^2}/{k_{\theta}}-
\sqrt{(w+2\alpha k_1)^2+k_2^2}$, leading to a van Hove singularity at $\mathcal{E}_{sp}(w,0)=k_{\theta}/2-w(1+\alpha)$. This
is similar to numerical results found in Ref.\ \onlinecite{Santos2}.

There are two other saddle points in the first BZ, near $M_b$ and $M_c$ in Fig. \ref{2BZ}.
Dispersions
for these can be obtained in a way very similar to that of $M_a$ by employing an appropriate
unitary transformation, shifting the zero of momentum for one of the two layers by
$\vec G_1$ or $\vec G_2$.  Up to 120$^\circ$ rotations, the resulting spectra are essentially
identical to that of $M_a$.

\emph{Magnetic Translation (MT) Operators} -- To incorporate a uniform perpendicular
magnetic field we introduce a vector potential $\vec A=B(-y/2,x/2)$.
To study the small $B$ limit it is convenient to work with momentum-space wavefunctions,
so that the momentum operators ${p}_i$
entering Eq.\ \ref{BasicModel}
are replaced by ${\Pi}_{1,2}=k_{1,2}\pm\frac{i}{2\ell_B^2}\partial_{k_{2,1}}$.
In the momentum representation, the interlayer tunneling terms are
$\hat{V}_i=\hat{t}_i \tau(\vec{G}_i)$
where $\vec{G}_0=0$, with momentum translation operators
$
\tau(\vec{G})=e^{G_x \partial_{k_1}+ G_y \partial_{k_2}}.
$

To exploit the translational symmetries of the problem we define MT operators
\bea
  T_1(G_x)=\exp{\left[G_x(\partial_{k_1}-2i\ell_B^2k_2)\right]}\:,\\
  T_2(G_y)=\exp{\left[G_y(\partial_{k_2}+2i\ell_B^2k_1)\right]}\:,\label{MMT}
\eea
which commute with $\Pi_{1,2}$.
The combinations $T(\vec G_{1,2})\equiv T_1(\pm G_x)T_2(G_y)$ moreover commute
with the {\it full} Hamiltonian, as well as with one another, if
\be
  4\ell^2_BG_xG_y=2\pi N\:,
\label{FluxCondition}
\ee
for any integer $N$.
We focus on magnetic fields satisfying this equality.
Note such fields have the form $B_N=\bar B/N$, so that our analysis applies to a
dense set of small magnetic fields.

Eigenfunctions of the Hamiltonian can also be expressed as eigenfunctions of
MT operators that commute with $H$, and it is convenient to choose the particular combination
$T(\vec{G}_1)T(\vec{G}_2) \equiv T^2_2(G_y)$ and $T(\vec{G}_1)$ for this purpose.
To see how this plays out, we consider spinor wavefunctions written in the form
$\vec{\psi}(k_1,k_2)=\int d{\tilde k}
e^{-2i\ell_B^2k_1k_2+4i\ell_B^2\tilde k k_2}\vec{\psi}'(k_1,{\tilde k})$.
In the absence of interlayer coupling, $\tilde k$ is a good quantum number
and eigenfunctions
of the Hamiltonian involve harmonic oscillator states whose centers lie near
$\tilde k$.  Thus $\tilde k$ can be viewed as a momentum-space guiding
center coordinate.  More generally,
the requirement that wavefunctions be eigenvectors of $T^2_2(G_y)$
dictates that $e^{8i\ell_B^2G_y\tilde k}$ be the
same for all the $\vec{\psi}'(k_1,{\tilde k})$'s entering a wavefunction.
The integral over $\tilde k$ then becomes a discrete sum.
To see the effect of interlayer
coupling one needs to notice that
the action of momentum shift operator $\tau(\vec{G}_1)$ appearing in the
interlayer coupling on $\vec\psi$ becomes
$$
\tau'(\vec{G}_1) \vec\psi'(k_1,\tilde k)=
e^{-2i\ell_b^2(k_1-2\tilde k)G_y}\vec\psi'(k_1+G_x,\tilde k +G_x/2).
$$
This is consistent with the allowed discrete values of $\tilde k$ for a given wavefunction
provided Eq.\ \ref{FluxCondition} is obeyed.

Thus $\vec{\psi}$ can be written as a sum over wavefunctions $\vec{\psi}'(k_1,\tilde k)$
with $\tilde k = \tilde k_0+\lbrace ...,-G_x/2,0,G_x/2,... \rbrace$ and
$0 \le \tilde k_0 < G_x/2$.  The set of $\tilde k$'s one must retain is further
reduced by use of a second MT symmetry condition, $T(\vec G_1)\vec\psi = e^{i \theta} \vec\psi.$
This becomes the condition
$e^{2i\ell_B^2G_xG_y+4i\ell^2_B G_y \tilde k} \vec \psi'(k_1+G_x,\tilde k+G_x)=
e^{i\theta} \vec \psi'(k_1,\tilde k)$.
Ultimately one needs to only compute \emph{two} functions, e.g., $\vec \psi'(k_1,\tilde k_0)$ and
$\vec \psi'(k_1,\tilde k_0+G_x/2)$.

Some comments are in order.  First, the reduction of the
wavefunction to two functions of $k_1$ was possible because of our
gauge choice \cite{footnote1}. Secondly, since $\vec
\psi'(k_1,\tilde k)$ involves a single continuous variable, $k_1$,
it can be approximated conveniently in a semiclassical approach.
Because we need to
retain two values of $\tilde k$, these wavefunctions must be
represented in two BZ's \cite{SCcase}.  Finally, while the two-BZ semiclassical
description is strictly valid only for fields satisfying Eq.\
\ref{FluxCondition}, we will treat $B$ as a continuous variable.
This captures the broad shape of the spectrum, but misses small gaps in
what turn out to be narrow bands in the low field limit \cite{Hofstadter}.

\emph{Semiclassical wavefunctions} -- Assuming $\ell_B$ is larger than any other
length scale in the problem (weak fields), we may use a
gradient expansion for the wavefunctions \cite{Chambers},
$\vec\psi'(k_1,\tilde k)\sim \exp\left[\ell_B^2S_{-1}+S_0+...\right]$.
We again start with uncoupled layers.  Defining
$q_y^{\pm}(k_1)=\Delta_y \pm Q_y(k_1)$, with
$Q_y(k_1)=\sqrt{E^2-(k_1-\Delta_x)^2}$,
and $\Delta_x=\tilde k$, $\Delta_y =(-)k_{\theta}/2$ for the top (bottom) layer,
the lowest non-trivial contribution has the form
\be
\vec\psi'_{\pm}\sim
  e^{i\ell_B^2\int^{k_1}dk_x q_y^{\pm}(k_x)}\:.
\label{semiclassical}
\ee
The (spinor) coefficient of the wavefunction is determined at higher order
in $1/\ell_B^2$ \cite{BerrySpinor}, and is not included in our analysis.
The set of momenta $\{(k_x,q^{\pm}_y(k_x))\}$
represent contours of constant energy above and below a Dirac point.
When $Q_y(k_1)$ approaches 0, these two curves approach one another,
and the semiclassical approximation breaks down.  To account for this
one employs matching conditions \cite{BenderOrszag} at each turning point.
These work simultaneously at certain discrete
energies, yielding a spectrum with spacing matching the exact result
for Landau levels of a single Dirac point Hamiltonian.

This result is essentially correct even in the presence of interlayer
tunneling when one considers levels close in energy to that of the Dirac points.
For energies near those of the saddle points, one must develop further connection
formulae among the different
semiclassical trajectories \cite{Chambers}.  This is most easily
implemented for $\hat{V}_0$, which connects trajectories near
$M_a$ in Fig.\ \ref{2BZ}.  The cases of $\hat V_{1,2}^{(\dag)}$ are somewhat more complicated \cite{unpub}.
$\hat V_{1}$ connects the wavefunction for $\tilde k$ in the top layer with
with the bottom layer for $\tilde k-G_x/2$ through the saddle point $M_c$
via the operator $\tau'(\vec G_1)$. The problem becomes closely analogous
to that of the $M_a$ saddle point if one applies a unitary transformation,
shifting the bottom component of the wavefunctions by $\tau'(-\vec G_1)$.
This is represented conveniently by placing one
quarter of the BZ for $\tilde k-G_x/2$ continuously onto the upper right
side of the $\tilde k$ BZ.  Similar constructions for $\hat V_{1}^{\dag}$,
$\hat V_{2}$, and $\hat V_{2}^{\dag}$ bring in another quarter of the $\tilde k-G_x/2$
BZ on the lower right, and half of the $\tilde k + G_x/2$ BZ on the left,
yielding a doubled BZ in the form of a rectangle.  This is illustrated
in Fig.\ \ref{2BZ} along with relevant semiclassical orbits, which are
labeled with unprimed (primed) numbers for the $\tilde k$ ($\tilde k \pm G_x/2$)
BZ.



Wavefunctions for the full system involve amplitudes multiplying functions of the form
in Eq.\ \ref{semiclassical}, with the caveat that $\vec\Delta$ represents the location of the
Dirac point around which an orbit is centered.  We assign an amplitude for each trajectory
that enters or exits a saddle point, which are related to one another in several ways.
({\it i}) Each trajectory has an amplitude ${a}_i^{\bullet}$ to exit from some saddle
point and an amplitude ${a}_i^{\circ}$ to enter another.  These are related
by ${a}_i^{\bullet}=(1,\pm i)e^{i\Phi_i}{a}_i^{\circ}$,
where $\Phi_i/\ell_B^2$ is the area between the trajectory (which begins and ends at the
points of closest approach to the saddle points) and the $q_y=0$ axis in Fig.\ \ref{2BZ}.  This area is
taken to be positive (negative) if the trajectory is above $k_1$-axis and moves to the right (left).
Factors of $\pm i$ must be inserted if
there is a left or right turning point in the trajectory \cite{BenderOrszag}. {\it (ii)} At each saddle point shown in Fig.\ \ref{2BZ},
there are two incoming trajectories and two outgoing ones.  These are related by
an $S$-matrix, which we discuss in more detail below.
{\it (iii)} Trajectories exiting the doubled
BZ on the left or right are related to ones entering on the opposite side
due to the periodicity imposed by the MT operators.  The effect of this can be
incorporated in the matrices relating different amplitudes
with some added (energy independent) phase factors \cite{details}.
In practice, their presence only impacts the spectrum for energies rather close
to that of the saddle points.

The $S$-matrix associated with the saddle points can be obtained
through the two-band model, Eq.\ (\ref{2band}). Introducing the
magnetic field by adding a vector potential to the momentum ${\bf
k}$, one finds the eigenvalue equation can be reduced to a single
component problem in the neighborhood of the saddle point at
$k_1=w$. With a gauge transformation to Landau gauge, one obtains
an eigenvalue equation involving a massive particle in an inverted
parabolic potential,
$\left[\frac{d^2}{dX^2}+\epsilon+\frac{X^2}{4}\right]\psi=0$ with
$\epsilon=\frac{\ell_B^2(E'^2-w^2)}{2\sqrt{(w-E')/k_{\theta}}}$
and
$X\equiv\sqrt{2}(k_1-w)\ell_B\left[\frac{w-E'}{k_{\theta}}\right]^{\frac{1}{4}}$ \cite{details}.
Here we define $E'=E-k_{\theta}/2$. The resultant $S$-matrix is
then obtained by standard methods \cite{Connor,Herb1}, yielding
\be
  S_0=\frac{e^{-i\Psi}}{\sqrt{1+e^{-\pi\epsilon}}}\left(\begin{array}{cccc}
     1 & ie^{-\pi\epsilon/2}\\
    ie^{-\pi\epsilon/2} & 1
    \end{array}\right)\:,\label{Smatrix}
\ee with
$\Psi=\epsilon+\arg\Gamma(\frac{1}{2}+i\epsilon)-\epsilon\ln{|\epsilon|}$.
Eq.\ (\ref{Smatrix}) suggests that $S_0\rightarrow \hat{\mathbb
I}_2$ for $E'\ll -w$ and $S_0\rightarrow i\hat\sigma_x$ for $E'\gg
-w$.  These two limits define intra- and inter-layer dominant
scattering regimes, respectively. As written, $S_0$ applies
directly to $M_a$; for $M_a'$, $M_b(^{\prime})$ and $M_c(^{\prime})$, certain matrix
elements are multiplied by phase factors related to the
eigenvalues of $T_2^2(G_y)$ and $T(\vec G_1)$ \cite{details}.
These only have noticeable affect quite close to the
saddle point energy.

The description above yields 24 independent amplitudes and 24
equations relating them. It is convenient to group these
amplitudes into the six four-component columns,
$(2,4,2',4')^T_{(\bullet,\circ)}$,
$(1,6,1',6')^T_{(\bullet,\circ)}$, and
$(3,5,3',5')^T_{(\bullet,\circ)}$. The labels for portions of the
trajectories are displayed in Fig.\ \ref{2BZ}. Requiring single-valued wavefunctions then leads to the condition \cite{details}
\be
  \rm det\left[\hat\mathcal Q_a\hat\mathcal R_{ab}\hat\mathcal Q_b\hat\mathcal R_{bc}\hat\mathcal Q_c\hat\mathcal R_{ca}- \hat\mathbb I_4\right]=0\:.\label{final_expression}
\ee In this representation, the $\hat{\mathcal Q}$'s are
4 $\times$ 4 matrices which treat scattering
through the $M_{i}$ and $M_i'$ SP's together
($i=a,b,c$).
The unitary matrix $\hat{\mathcal
R}_{ij}=\exp[-i\ell_B^2(\mathcal A_{ij}\hat{\mathbb I}_4-\mathcal
D_{ij}\hat\Gamma)]$ with $\hat{\Gamma}=\hat\sigma_z\otimes
\hat{\mathbb I}_2$ encodes areas swept out by electron orbits between
the $i$th and $j$th SP. Precise definitions of the
$\hat{\mathcal Q}$'s, $\mathcal A$'s, and $\mathcal D$'s are given
in Ref.\ \onlinecite{details}.

\begin{figure}
\input{epsf}
\includegraphics[width=0.5\textwidth]{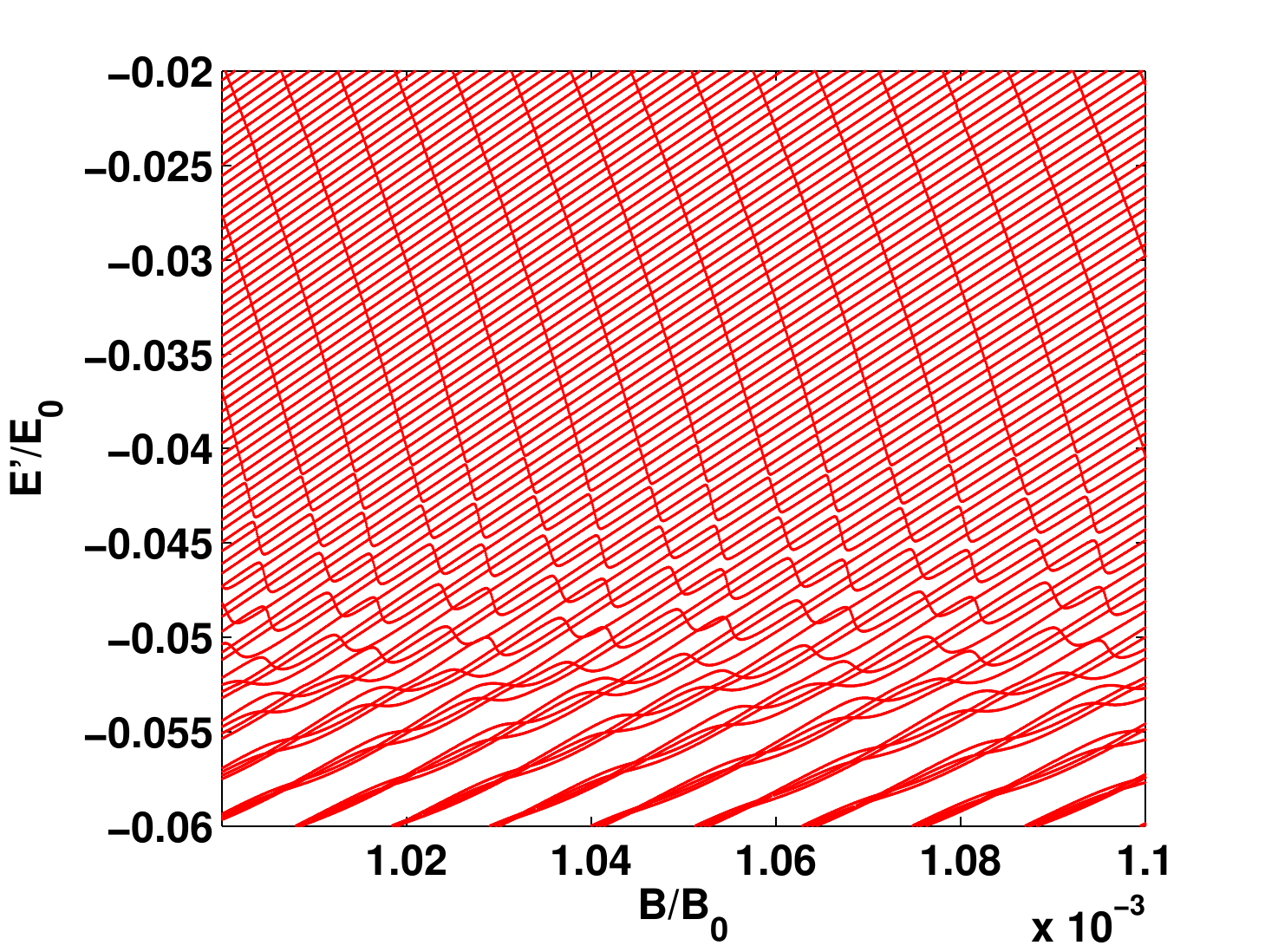}
\caption{Energy spectrum near the SP energy for the interlayer hopping $\alpha=0.05$.
Here the reference of energy corresponds to zero of $E'\equiv E-v_Fk_{\theta}/2$.
Energy units are $E_0=v_Fk_{\theta}$, magnetic field units are $B_0=h c k^2_{\theta}/e$. 
}\label{LL}
\end{figure}

Numerical solutions to the problem described above are
illustrated in Fig.\ \ref{LL}. and are consistent with direct
numerical diagonalization \cite{Bistritzer2} of the Hamiltonian in
Eq. \ref{BasicModel} \cite{unpub}.  Relatively simple behavior is
apparent well below and above the saddle point energy, which may
be understood analytically. Below the saddle point, interlayer
tunneling is negligible, leading to $\hat{\mathcal Q}_i
\rightarrow \hat{\mathbb I}_4$. Moreover, $\sum_\gamma \mathcal
A_\gamma = \mathcal A=\pi E^2$, which is the area enclosed by a
trajectory in an uncoupled layer, and $\sum_{\gamma}\mathcal D_{\gamma}=0$.
This leads to the standard Dirac-Landau level spacing.

Above the saddle point, one finds \cite{details}
$\hat{\mathcal Q}_a \rightarrow i \hat{\mathbb I}_2\otimes\hat\sigma_x$ and $\hat{\mathcal Q}_b=\hat{\mathcal Q}_c \rightarrow i\hat\sigma_x\otimes\hat\sigma_x$. These anticommute with $\hat\Gamma$, and one can show that Eq. \ref{final_expression} is satisfied if $e^{i\ell_B^2(\mathcal{A\pm X})}+i=0$ for either one of the two signs in the exponent.
In this expression, $\mathcal X = \mathcal D_{ab}+\mathcal D_{ca}-\mathcal D_{bc}=3\sqrt{3}k_{\theta}E/2$, and $\mathcal X(E=\frac{k_{\theta}}{2})$ corresponds to the area of half of a single BZ.
The quantities $\mathcal A +(-) \mathcal X$ are areas related to
the star-like orbits, which increase (decrease) in magnitude with energy,
leading to coexisting electron- and hole-like levels.

\emph{Discussion} -- The spectrum predicted above resolves some
apparent inconsistencies among recent results.  Studies which
include only one saddle point \cite{deGail1,Choi,Azbel,Herb2} yield
purely electron-like spectra. By contrast, one expects hole-like
orbits surrounding local maxima to come down towards the saddle
point, as shown in tight-binding studies of the twisted
bilayer \cite{Moon}. These pictures are in a sense both
correct.  When several SP's are degenerate in energy, the
necessity to include multiple BZ's allows electron- and hole-like
orbits to coexist. Importantly, this structure explains how levels
rising from below the SP and levels falling from above with
increasing $B$ evolve: the levels anti-cross, and all ultimately
move to high energy when the field is sufficiently large, as is
evident in Fig.\ \ref{LL}. This behavior should appear in many
systems where degenerate, distinguishable SP's allow a transition
between topologically distinct semiclassical orbits, including
graphene on boron nitride
substrates \cite{Hofstadter_exp1,Hofstadter_exp2}, and in single
layer graphene at high energy \cite{levitov}.  This behavior is
also apparent in the surface states of crystalline topological
insulators in a magnetic field \cite{Okada}.

The peculiar Landau level structure and the associated
semiclassical orbits in our model should have a number of
experimental ramifications. For sufficiently clean samples, the
level structure itself could be detected directly in
tunneling \cite{Andrei}.  Cyclotron resonance \cite{CR1,CR2,CR3} brings
another interesting perspective: since star-like orbits tunnel
periodically between layers, electromagnetic waves with electric
field perpendicular to the layers should couple to them and allow
absorption, whereas in truly two-dimensional systems this would
not be possible. (Preliminary calculations \cite{unpub} demonstrate
that this is indeed the case.) Thermodynamically, converging
hole-like and electron-like orbits at the saddle point energy
should lead to cusp-like behavior in magnetic
susceptibility \cite{Vignale}.  Finally, breaking the symmetry
among the saddle points, for example by strain or a periodic
potential \cite{Luis}, can in principle induce open orbits, which
might be observed in transport as a metal-insulator transition.

\emph{Acknowledgements} --  The authors are grateful to Luis Brey
and Pablo San Jose for useful discussions, and to S.-X. Zhang for
pointing out Ref.\ \onlinecite{Okada}.  This work was supported by
the NSF through Grant No. DMR-1005035, and the US-Israel
Binational Science Foundation, through Grant No. 2008256.


\pagebreak

\onecolumngrid

\section*{Supplementary Material}

\subsection{{\bf Saddle Point Dispersion from $ \vec k \cdot \vec p $ Approximation}}

Near $M_a$, the electron wavefunction may be taken as proportional to $e^{i\vec k\cdot \vec r}$, with small $k=\sqrt{k_{1}^2+k_2^2}$, independent of layer index. Only
$\hat V_0$ is relevant to the spectrum in this small range of momentum, and small terms of
order $k$ can be treated perturbatively. The form of $\hat V_0$ is simplified by the transformation
$\hat U^{\dag}\hat{t}_0 \hat U=\hat{\mathbb I}_2+\hat\sigma_z$ with
$\hat U=\exp{[-i\frac{\pi}{4}\hat\sigma_y]}$, and the transformed $\hat V_0$
now has only one nonzero matrix
element. With the unitary transformation,

\be
  B=\left(\begin{array}{cccc}
    \hat U & 0 \\
    0 & \hat U
  \end{array}\right)\times
  \left(\begin{array}{cccc}
    1 &  &  \\
    & \hat \sigma_x & \\
     & & 1
    \end{array}\right)\:, \label{Btransform}
\ee in which the latter matrix exchanges the second and third rows
and columns, we may transform the Hamiltonian into

\be
  \mathcal H_{sp}=B^{\dag}HB=\left(\begin{array}{cccc}
    2w\hat \sigma_x & i\frac{k_{\theta}}{2}\hat \sigma_z \\
    -i\frac{k_{\theta}}{2}\hat \sigma_z & 0
    \end{array}\right)+
    \left(\begin{array}{cccc}
    k_1\hat \mathbb I_2 & -ik_2\hat \mathbb I_2 \\
    ik_2\hat \mathbb I_2 &  -k_1\hat \mathbb I_2
    \end{array}\right)\:.\label{Hsp}
\ee It is easy to diagonalize the first matrix, $H_0$, and the its
four eigenvalues are given by

\be
  \lambda^{+}_{1,2}=-\lambda^{-}_{1,2}=\frac{\sqrt{k_{\theta}^2+4w^2}}{2}\pm w\:,
\ee where $w$ is positive and we assume $\lambda^{+}_1>\lambda^{+}_2$, $\lambda^{+}_1=-\lambda^{-}_1$.
Defining ket vectors in terms of eigenstates of $\hat\sigma_x$,
$\sigma_x|\pm>=\pm|\pm>$ and the constant $\beta=\sqrt{\lambda^+_2/\lambda^+_1}$, the normalized eigenvectors are given by

\be
  |\lambda^{+}_1\rangle=\frac{1}{\sqrt{2-4\alpha}}
  \left(\begin{array}{cccc}
    |+>\\
    -i\beta|->
    \end{array}\right)\label{eigenstate1}\:,
\ee and

\be
  |\lambda^{+}_2\rangle=\frac{1}{\sqrt{2+4\alpha}}
  \left(\begin{array}{cccc}
    |->\\
    -i\beta^{-1}|+>
    \end{array}\right)\:,\label{eigenstate2}
\ee for the positive energy eigenstates. Note that $\beta^2=1-4\alpha$. Their negative energy counterparts are

\be
  |\lambda^{-}_1\rangle=\frac{1}{\sqrt{2-4\alpha}}
  \left(\begin{array}{cccc}
    |->\\
    i\beta|+>
    \end{array}\right)\label{eigenstate_m1}\:,
\ee and

\be
  |\lambda^{-}_2\rangle=\frac{1}{\sqrt{2+4\alpha}}
  \left(\begin{array}{cccc}
    |+>\\
    i\beta^{-1}|->
    \end{array}\right)\:.\label{eigenstate_m2}
\ee Because $\lambda^{+}_i=-\lambda^{-}_i$ for $i=1,2$, there is an antiunitary relation between the states of the form

\be
  |\lambda^{+}_i>=\left[I_2\otimes\sigma_3\right]\mathcal K\ |\lambda^{-}_i>\:,\ i=1,2
\ee in which complex conjugation is represented by $\mathcal K$.

The perturbation in $k_1$ and $k_2$, the second term in Eq.\ \ref{Hsp}, may be expressed as

\be
  H_1=k_1[\hat\sigma_3\otimes \hat{\mathbb I}_2]+k_2[\hat\sigma_2\otimes \hat{\mathbb I}_2]\:,
\ee with which one may verify the matrix elements

\be
  <\lambda^{+}_1|H_1|\lambda^{+}_1>=-<\lambda^{+}_2|H_1|\lambda^{+}_2>=2\alpha k_1\:.\label{diagonal}
\ee These diagonal matrix elements contain no contribution from $k_2$. On the other hand, $k_2$ does appear in the off-diagonal matrix element,

\be
  <\lambda^{+}_1|H_1|\lambda^{+}_2>=-k_2\:.
\ee These matrix elements define an approximate projection of
$H_1$ into the positive eigenvalue subspace of the $k=0$
Hamiltonian in Eq.\ \ref{Hsp}.  A correction to this can be
included in the diagonal elements in Eq.\ (\ref{diagonal}) using
second-order perturbation theory, with the  negative energy
states, Eqs.\ \ref{eigenstate_m1} and \ref{eigenstate_m2}, being
the intermediate states. Using

\be
   <\lambda^{+}_1|H_1|\lambda^{-}_1>=0\:,\ <\lambda^{+}_1|H_1|\lambda^{-}_2>=k_1\:
\ee and

\be
  <\lambda^{+}_2|H_1|\lambda^{-}_2>=0\:,\ <\lambda^{+}_2|H_1|\lambda^{-}_1>=k_1\:,
\ee the correction to the both diagonal terms is the same, with

\be
  \frac{|\langle\lambda^{+}_1|\mathcal{H}_1|\lambda^{-}_2\rangle|^2}{\lambda^{+}_1-\lambda^{-}_2}=
  \frac{|\langle\lambda^{+}_2|\mathcal{H}_1|\lambda^{-}_1\rangle|^2}{\lambda^{+}_2-\lambda^{-}_1}
  =\frac{k_1^2}{k_{\theta}}\:.
\ee Putting these results together, the projection of $H_1$ onto
the states $\{|\lambda^+_{1,2}>\}$ can be expressed approximately
as a two-band Hamiltonian,

\be
  H_1\mapsto\frac{k_1^2}{k_{\theta}} \hat{\mathbb I}_2 + 2\alpha k_1\hat\sigma_z -k_2\hat\sigma_x\:.
\ee Together with the unperturbed Hamiltonian
$H_0 \mapsto \frac{\lambda_1^++\lambda_2^+}{2}\hat \mathbb I_2 +
\frac{\lambda_1^+-\lambda_2^+}{2}\hat\sigma_3$, the two-band
Hamiltonian in the main text (Eq.\ 2) is obtained.

\subsection{Saddle Point Hamiltonian from Two Band Model}

For the purpose of computing the $S$-matrix, we may choose any convenient gauge.  The wavefunctions
well away from the saddle point have distinct in-coming and out-going characters on either side
of it, so that a gauge transformation does not affect the $S$-matrix itself.  To compute the $S$-matrix
we adopt Landau gauge, so that introducing the vector potential can be implemented via the substitution $k_2\rightarrow k_2-\frac{i}{\ell_B^2}\partial_{k_1}$, while $k_1$ remains unchanged. The energy reference is set by $E'=E-k_{\theta}/2=0$. The corresponding equations for the two-band model become

\bea
  (V_A-E')u+\frac{i}{\ell_B^2}\partial_{k_1}v&=&0\:,\\
  \frac{i}{\ell_B^2}\partial_{k_1}u+(V_B-E')v&=&0\:,
\eea where

\be
  V_{A(B)}(k_1)=\frac{(k_1\pm w)^2}{k_{\theta}}\pm(1\mp\alpha)w\:.
\ee
One may eliminate the $u$ term to arrive at

\be
  \frac{1}{\ell_B^4}\frac{V_A'}{V_A-E'}v'-(V_A-E')(V_B-E')v-\frac{1}{\ell_B^4}\frac{\partial^2}{\partial k_1^2}v=0\:,
\ee
and furthermore eliminate the derivative term $v'$ by  writing $v=\sqrt{V_A-E'}\psi$ to obtain

\be
  -\frac{1}{\ell_B^4}\psi''+(E'-V_A)(V_B-E')\psi=0+O(\ell_B^{-4})\:.\label{Webber}
\ee
For the band containing the saddle point, $E'<0$.  The factor $(V_A-E')(V_B-E')$ on the left-hand side of the above equation in this situation has the form of an inverted parabola in the neighborhood of
$k_1=w$, which can be approximated as

\be
(V_A-E')(V_B-E') \approx 
[E'-(1+3\alpha)w]\left[\frac{(k_1-w)^2}{k_{\theta}}-w(1+\alpha)-E'\right]\:.
\ee

Then we may then rewrite Eq.\ (\ref{Webber}) in the form

\be
  \left[\frac{d^2}{dX^2}+\frac{X^2}{4}+\epsilon\right]\psi=0\:,
\ee with

\be
  X=\sqrt{2}\ell_B(k_1-w)\left[\frac{(1+3\alpha)w-E'}{k_{\theta}}\right]^{1/4}\:
\ee and

\be
  \epsilon=\sqrt{k_{\theta}}\ell_B^2\frac{[E'+(1+\alpha)w][E'-(1+3\alpha)w]}{2\sqrt{(1+3\alpha)w-E'}}.
\ee From this expression one may compute the $S$-matrix using
standard methods as described in the text.  Incoming and outgoing
states on either side of $X=0$ correspond to such states for the
original Hamiltonian, and this same $S$-matrix connects the
amplitudes for those states. The expressions for $X$ and
$\epsilon$ in the main text are obtained by setting $\alpha=0$ for
the purpose of simplifying the expressions; the actual numerical
computation still uses the expressions listed here.

\vfill\break

\subsection{Derivation of Equation 8}



The incoming and outgoing amplitudes near saddle points $M_a$ and $M_a'$ are related through,

\be
  \left(\begin{array}{cccc}
  2\\
  4\\
  2'\\
  4'\end{array}\right)_{\bullet}=
\left(\begin{array}{cccc}
    S_0 & 0   \\
    0 & U^{\dag}_{\phi}S_0V_{\phi}
    \end{array}\right)
\left(\begin{array}{cccc}
  1\\
  6\\
  1'\\
  6'\end{array}\right)_{\circ}\equiv\mathcal Q_a
\left(\begin{array}{cccc}
  1\\
  6\\
  1'\\
  6'\end{array}\right)_{\circ}\:,
  \label{prop1}
\ee in which the parameter $\phi$ encodes the boundary condition
between the edges of the doubled BZ, and the basic $S$-matrix
$S_0$ is given in the main text.  The 2x2 unitary matrices,

\be
  U_{\chi}\equiv\left(\begin{array}{cccc}
    1 & 0   \\
    0 & e^{i\chi}\end{array}\right)\:,\
  V_{\chi}\equiv\left(\begin{array}{cccc}
    e^{i\chi} & 0   \\
    0 & 1 \end{array}\right)\:,
\ee specify the boundary conditions.

\begin{figure}
\input{epsf}
\includegraphics[width=0.65\textwidth]{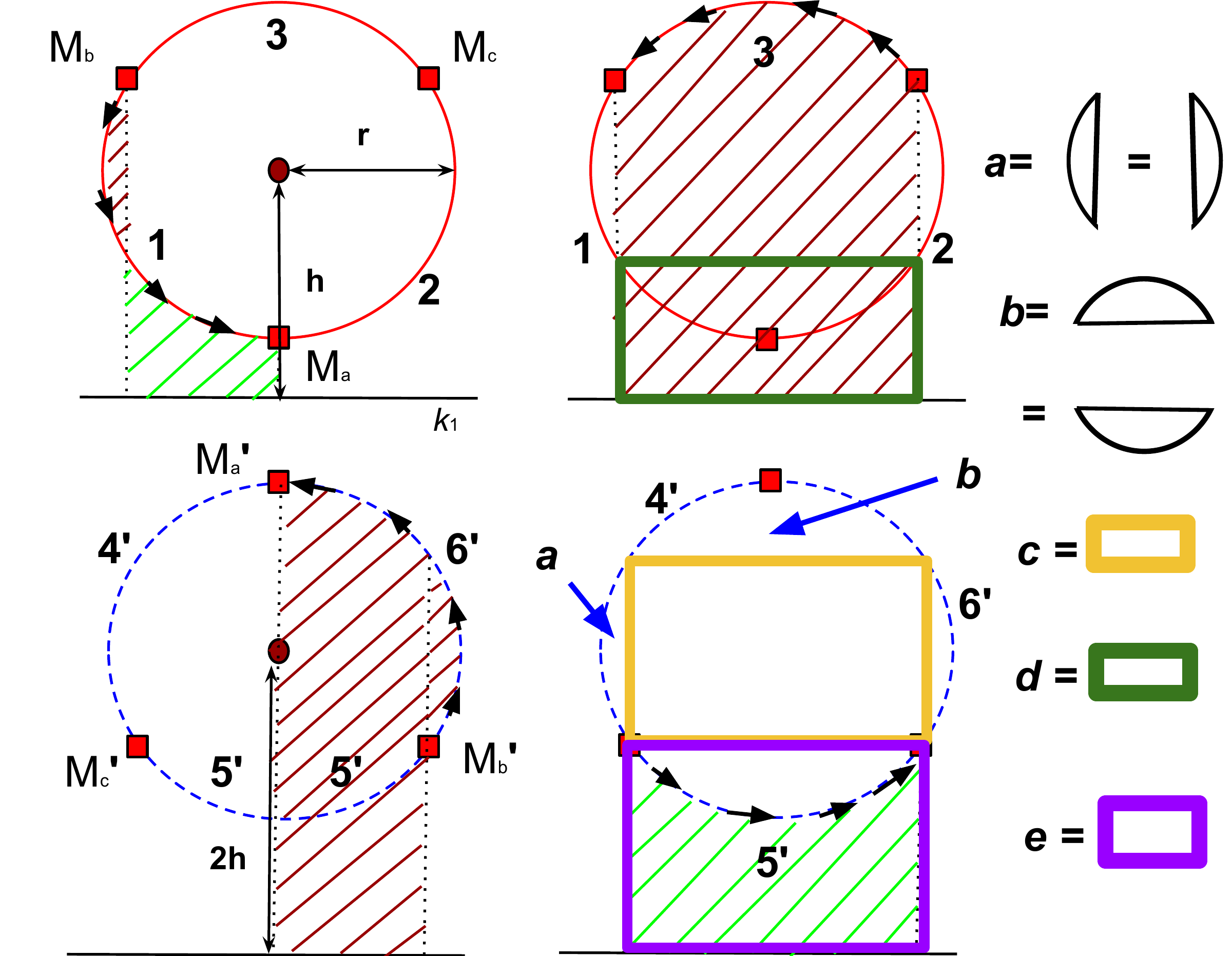}
\caption{(Color online)
Representation of $\mathcal A$'s and $\mathcal D$'s in Eqs.\ \ref{prop2}, \ref{prop4} and \ref{prop6} by the
shaded areas associated with the representative arcs. Circles in top row are the trajectories of top layer in first BZ, while those in bottom row are for the trajectories of bottom layer in second BZ (See Fig.\ 1 in main text). Because the orbits are symmetric about the $k_1$ axis, arc 1 in top-left is the representative of arcs $\{1,2,4,6\}$ all of which under the specified circulation correspond to the same area listed in Table \ref{area_table}. Similarly, arc 3 on top-right, arc $6'$ on bottom left, and arc $5'$ on bottom right are the representative ones. The five elementary areas $a-e$ specified by either shape or color are listed in the right column. Referring to the circle in bottom-right, the circle of area $\pi r^2$ with $r=E$ is divided into three distinct parts, the gold box in the middle of area $c$, the side of area $a$ and the top/bottom portion of area $b$. The green box of area $d$ in top-right and the purple box of area $e$ in bottom-right are different because the Dirac points (the center of circle) in top and bottom rows have different distances, $h=k_{\theta}/2$ and $2h$, respectively, from the $k_1$ axis.}\label{BZBox}
\end{figure}

The subsequent phase accumulation between saddle points $M_a$ and $M_b'$ (arc 1 and 6 in first BZ) and that between $M_a'$ and $M_b$ (arc $1'$ and $6'$ in the second BZ) is represented by

\be
\left(\begin{array}{cccc}
  1\\
  6\\
  1'\\
  6'\end{array}\right)_{\circ}=e^{-i\mathcal A_{ab}}e^{i\mathcal D_{ab}\hat\Gamma}
\left(\begin{array}{cccc}
  1\\
  6\\
  1'\\
  6'\end{array}\right)_{\bullet}\equiv\mathcal R_{ab}
\left(\begin{array}{cccc}
  1\\
  6\\
  1'\\
  6'\end{array}\right)_{\bullet}\:,\label{prop2}
\ee in which $\hat \Gamma={\hat\sigma_z}\otimes{\hat\mathbb I_2}$.
The quantities $\mathcal A_{ab}$ and $\mathcal D_{ab}$ appearing
in the exponent combine to give the areas between the numbered
arcs and the $k_1$ axis. We shall represent these areas in terms
of the five elementary areas $a-e$ defined in Fig.\ \ref{BZBox}.
Before we proceed to show how the representation of area is done,
one should notice that arcs 1 and 6 (see Fig.\ 1 in the main text)
in first BZ should sweep out identical areas since the orbits are
symmetric about the $k_1$ axis and move in opposite directions. The same
is true for arcs $1'$ and $6'$ in the second BZ. 
One may show that

$$-(\mathcal A_{ab}-\mathcal D_{ab})=a+\frac{b-d}{2}$$ is the shading area
associated with arc 1 in the top-left of Fig.\ \ref{BZBox}. It can
be seen that $(-a)$ is the negative (brown) area contributed from
the left 
side of circle, and
$(d-b)/2$ is the positive (green) shaded area. Similarly, one may
show that the area associated with arc $6'$ is
$$-(\mathcal A_{ab}+\mathcal D_{ab})=a+\frac{b+c+e}{2}\:.$$
Note that
the overall minus sign is due to the choice of circulation of
those closed trajectories specified by the arrow in Fig.\
\ref{BZBox}. Continuing the same procedure, one can write down the
relations at the saddle point $M_b$ and $M_b'$,


\be
\left(\begin{array}{cccc}
  1\\
  6\\
  1'\\
  6'\end{array}\right)_{\bullet}= \hat Y\left(\begin{array}{cccc}
    U_{\theta}S_0U^{\dag}_{\theta} & 0   \\
    0 & V^{\dag}_{\theta}S_0V_{\theta}
    \end{array}\right)\hat Y
\left(\begin{array}{cccc}
  3\\
  5\\
  3'\\
  5'\end{array}\right)_{\circ}\equiv\mathcal Q_b
\left(\begin{array}{cccc}
  3\\
  5\\
  3'\\
  5'\end{array}\right)_{\circ}\:,
  \label{prop3}
\ee with

\be
  \hat Y=\left(\begin{array}{cccc}
    1 & 0 & 0 & 0  \\
    0 & 0 & 0 & 1 \\
    0 & 0 & 1 & 0 \\
    0 & 1 & 0 & 0
    \end{array}\right)\:.
\ee 
The reordering matrix $\hat Y$ implements the property that the $M_b$ and $M_b'$ SP's scatter
trajectories between different BZ's. $\theta$ is another phase angle parameter encoding the eigenvalues
under the MT operators. The next phase accumulation is given by

\be
\left(\begin{array}{cccc}
  3\\
  5\\
  3'\\
  5'\end{array}\right)_{\circ}=e^{-i\mathcal A_{bc}}e^{i\mathcal D_{bc}\hat\Gamma}
\left(\begin{array}{cccc}
    1 & 0   \\
    0 & V^{\dag}_{\phi}U_{\phi}\end{array}\right)
\left(\begin{array}{cccc}
  3\\
  5\\
  3'\\
  5'\end{array}\right)_{\bullet}\equiv \mathcal R_{bc}
\left(\begin{array}{cccc}
  3\\
  5\\
  3'\\
  5'\end{array}\right)_{\bullet}\:,
  \label{prop4}
\ee with the area
$$-(\mathcal A_{bc}-\mathcal D_{bc})=b+c+d$$ specifying the phase along arcs 3 and 5 (top-right in Fig.\ \ref{BZBox}), and $$-(\mathcal A_{bc}+\mathcal D_{bc})=b-e$$ specifying the phase along arcs $3'$ and $5'$ (bottom-right in Fig.\ \ref{BZBox}).
Finally, the scattering at $M_c$ and $M_c'$ may be written as

\be
\left(\begin{array}{cccc}
  3\\
  5\\
  3'\\
  5'\end{array}\right)_{\bullet}=Y\left(\begin{array}{cccc}
    V^{\dag}_{\theta}SV_{\theta} & 0   \\
    0 & U_{\theta}SU^{\dag}_{\theta}
    \end{array}\right)Y
\left(\begin{array}{cccc}
  2\\
  4\\
  2'\\
  4'\end{array}\right)_{\circ}\equiv \mathcal Q_c
\left(\begin{array}{cccc}
  2\\
  4\\
  2'\\
  4'\end{array}\right)_{\circ}\:,
  \label{prop5}
\ee and the subsequent phase accumulation by

\be
\left(\begin{array}{cccc}
  2\\
  4\\
  2'\\
  4'\end{array}\right)_{\circ}=e^{-i\mathcal A_{ca}}e^{i\mathcal D_{ca}\hat\Gamma}
\left(\begin{array}{cccc}
  2\\
  4\\
  2'\\
  4'\end{array}\right)_{\bullet}\equiv \mathcal R_{ca}
\left(\begin{array}{cccc}
  2\\
  4\\
  2'\\
  4'\end{array}\right)_{\bullet}\:.
  \label{prop6}
\ee The corresponding areas for arcs 2 and 4 are the same as those for arcs 1 and 6, and arcs $2'$ and $4'$ are the same as for arcs $1'$ and $6'$. Putting together Eqs. \ref{prop1}, \ref{prop2}, \ref{prop3}, \ref{prop4}, \ref{prop5} and \ref{prop6} leads to Eq. 8 in the main text.

Finally, for the next section it is useful to note the relations

\be
  \mathcal A_{ab}+\mathcal A_{bc}+\mathcal A_{ca}\equiv \mathcal A=2a+2b+c\:.
\ee
For energies below that of the saddle point, one finds (excluding corrections
of order $\alpha$)
 $\mathcal A =\pi E^2$,
which corresponds to the circular area associated with the trajectories of
energy below that of the saddle point. Moreover,

\be
  \mathcal D_{ab}+\mathcal D_{bc}+\mathcal D_{ca}=0\:,\ \mathcal D_{ab}-\mathcal D_{bc}+\mathcal D_{ab}=-(c+d+e)\equiv -\mathcal X\:.
\ee
Again excluding corrections of order $\alpha$, one finds $\mathcal X=3\sqrt{3}k_{\theta}E/2$. This is relevant for the quantization condition above the saddle point.


\subsection{Energy Level Conditions Above/Below Saddle Point}

For energy sufficiently below the saddle point, $E'\ll -w$, the
basic S-matrix reduces to $S_0 \mapsto \hat\mathbb I_2$, which
leads to all $\mathcal Q$'s equal identity matrix as well. Because
the sum of $\mathcal D$'s vanishes and the sum of $\mathcal A$'s equals the
Dirac circle area, it is easy to show that Eq.\ 8 in the text
reduces to

\be
  e^{i\ell_B^2\mathcal A}=1\:,
\ee which gives the ordinary Landau levels for single layer
graphene.

For energy sufficiently above that at the saddle point, the $S_0
\mapsto i\hat\sigma_1$. For simplicity, we set $\theta=\phi=0$.
The product of the six matrices $\hat\mathcal Q_a\hat\mathcal
R_{ab}\hat\mathcal Q_b\hat\mathcal R_{bc}\hat\mathcal
Q_c\hat\mathcal R_{ca}$ can be written as


\be
  (i)^3 e^{i{\mathcal A}}\
\left[\hat{\mathbb I}_2\otimes\hat\sigma_x\right]\
e^{-i\frac{\mathcal X}{4}\hat\Gamma}
\left[\hat\sigma_x\otimes\hat\sigma_x\right] e^{i\frac{\mathcal
X}{2}\hat\Gamma}
\left[\hat\sigma_x\otimes\hat\sigma_x\right] e^{-i\frac{\mathcal
X}{4}\hat\Gamma}=(-i)e^{i{\mathcal A}}\left[\hat{\mathbb
I}_2\otimes\hat\sigma_x\right]e^{-i\mathcal X\hat\Gamma}\:, \ee
where we have written $\mathcal D_{ab}=\mathcal D_{ca}=-2\mathcal
D_{bc}=-\mathcal X/4$. We have also used the facts that
$\hat\sigma_x\otimes\hat\sigma_x$ in the brackets
anticommutes with
$\hat\Gamma={\hat\sigma_z}\otimes{\hat\mathbb I_2}$, that its
square is the unit matrix . Eq. 8 in the text then reduces to


\be
   \det \left(\begin{array}{cccc}
    e^{i{\mathcal A}}+ie^{-i\mathcal X}\ \hat\sigma_x & 0   \\
    0 & e^{i{\mathcal A}}+ie^{i\mathcal X}\ \hat\sigma_x\end{array}\right)=0\:.
\ee which leads to two possible conditions for the allowed areas,

\be
  \cos\left[\ell_B^2(\mathcal A\pm\mathcal X)\right]=0\:.
\ee
Note that we have set $\ell_B^2=1$ in all expressions in this Supplement
except the
last one. The inclusion of boundary conditions specified by
$\theta$ and $\phi$ can be shown to yield identical spectra away
from the saddle point. However, for energies close to the saddle
point, the magnetic states are indeed altered by these parameters.
This is illustrated
in Fig.\ \ref{LL1} below, which show how the states in a range of energies
behave for various values of $(\theta,\phi)$.


\begin{table}[t]

\begin{tabular}{c | c c c c c  r }

arc number & area \\
 \hline
1,2& $-(a+\frac{b}{2})+\frac{d}{2}$ \\
4,6& $-(a+\frac{b}{2})+\frac{d}{2}$ \\
$1'$,$2'$& $-(a+\frac{b+c+e}{2})$\\
$4'$,$6'$& $-(a+\frac{b+c+e}{2})$\\
3,5 &  $-(b+c+d)$\\
$3'$,$5'$ & $-b+e$ 
\\
\end{tabular}

\caption[]{Representation of the area between the portions of trajectories (See Fig.\ 1 in main text for numbering) and $k_1$ axis using the five elementary areas shown in Fig.\ref{BZBox} of the supplement. The pair of numbered arcs appearing in the same row correspond to the same area due to the orbit symmetry with respect to $k_1$ axis and the fact that they move in opposite directions. Arcs 1, 3, $6'$, and $5'$ are four representative orbits for demonstrating the areas in terms of the elementary ones $a-e$ in Fig.\ \ref{BZBox}.
}\label{area_table}
\end{table}

\begin{figure}
\includegraphics[width=0.3\textwidth]{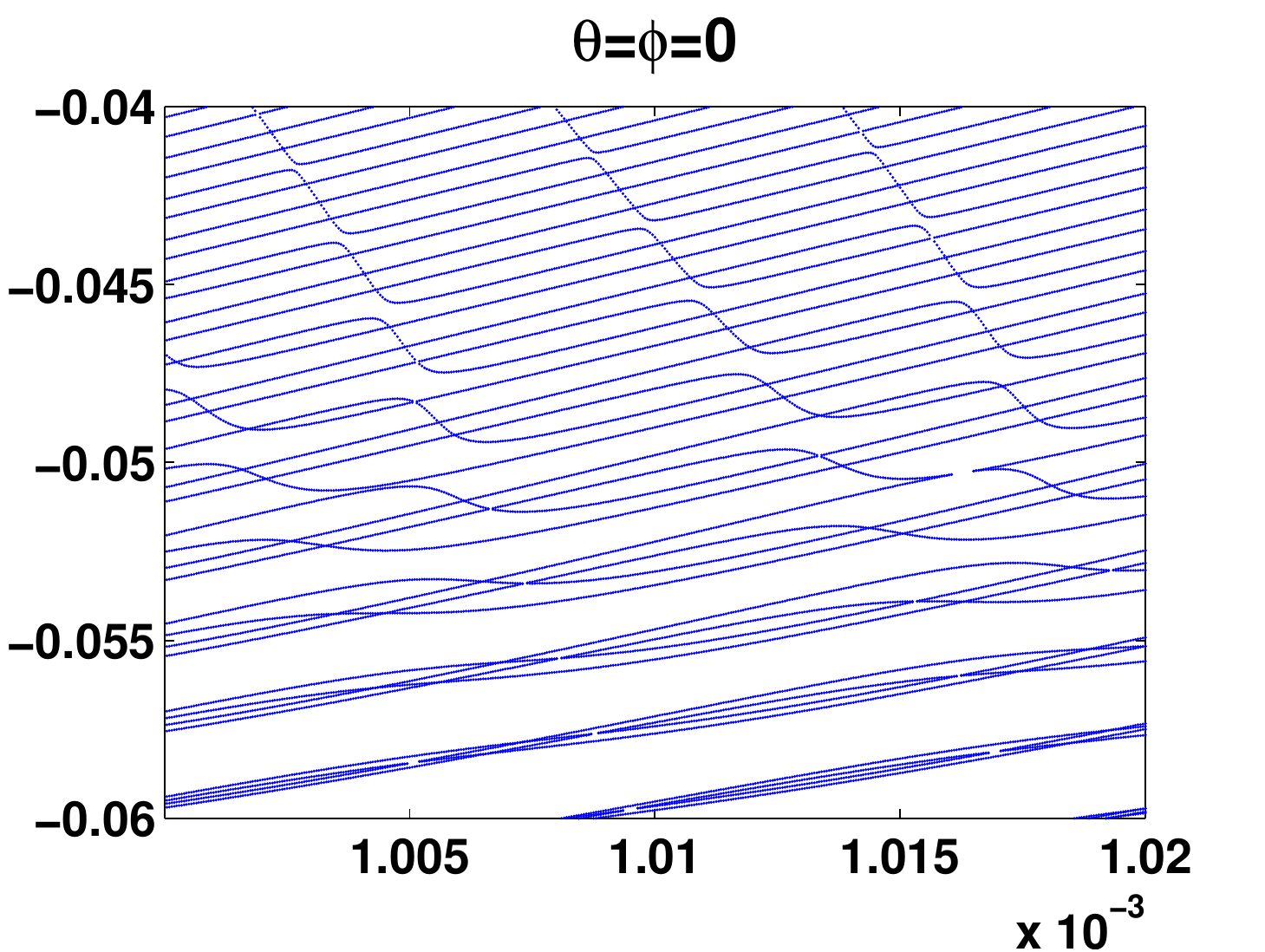}
\includegraphics[width=0.3\textwidth]{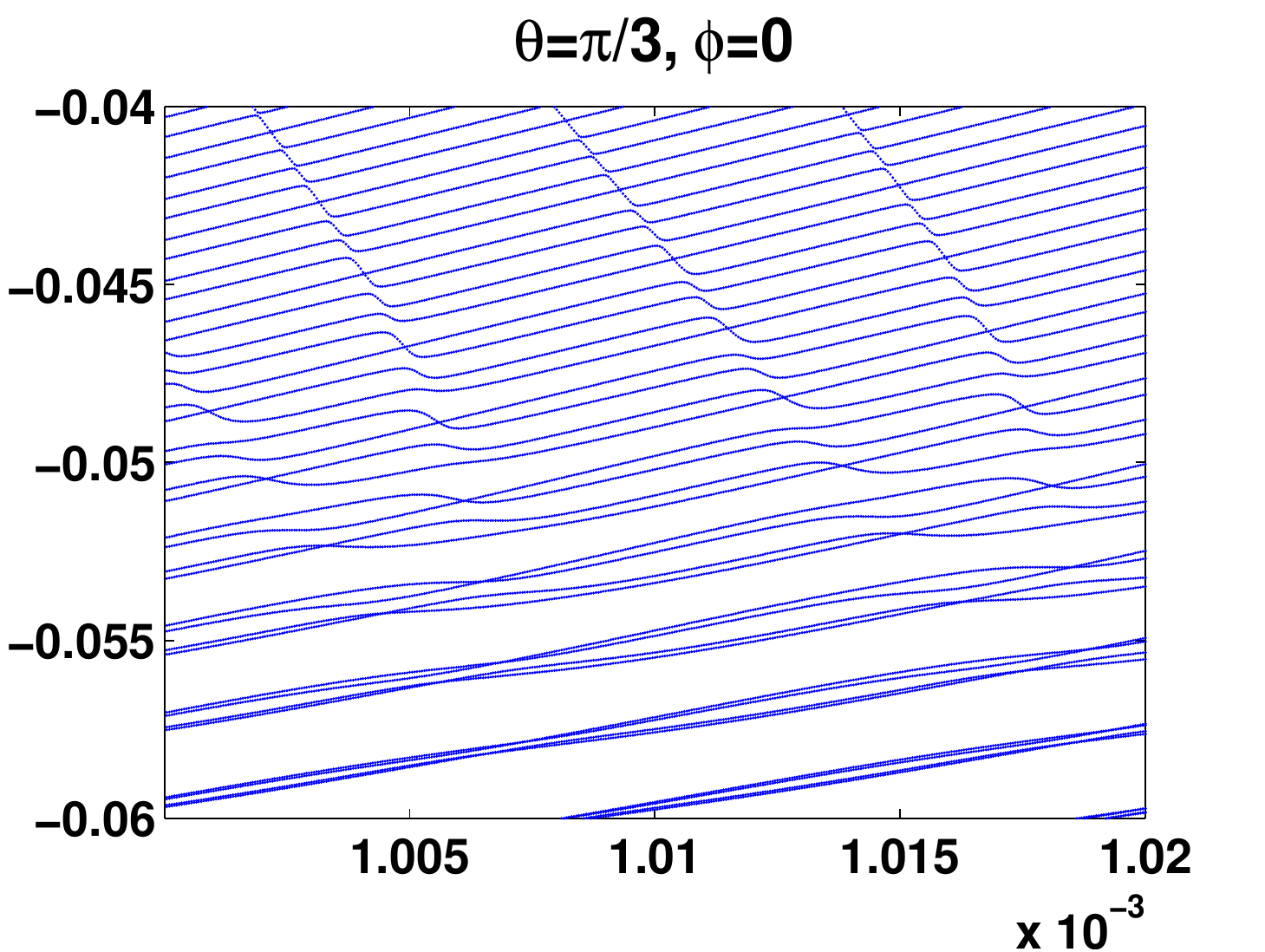}
\includegraphics[width=0.3\textwidth]{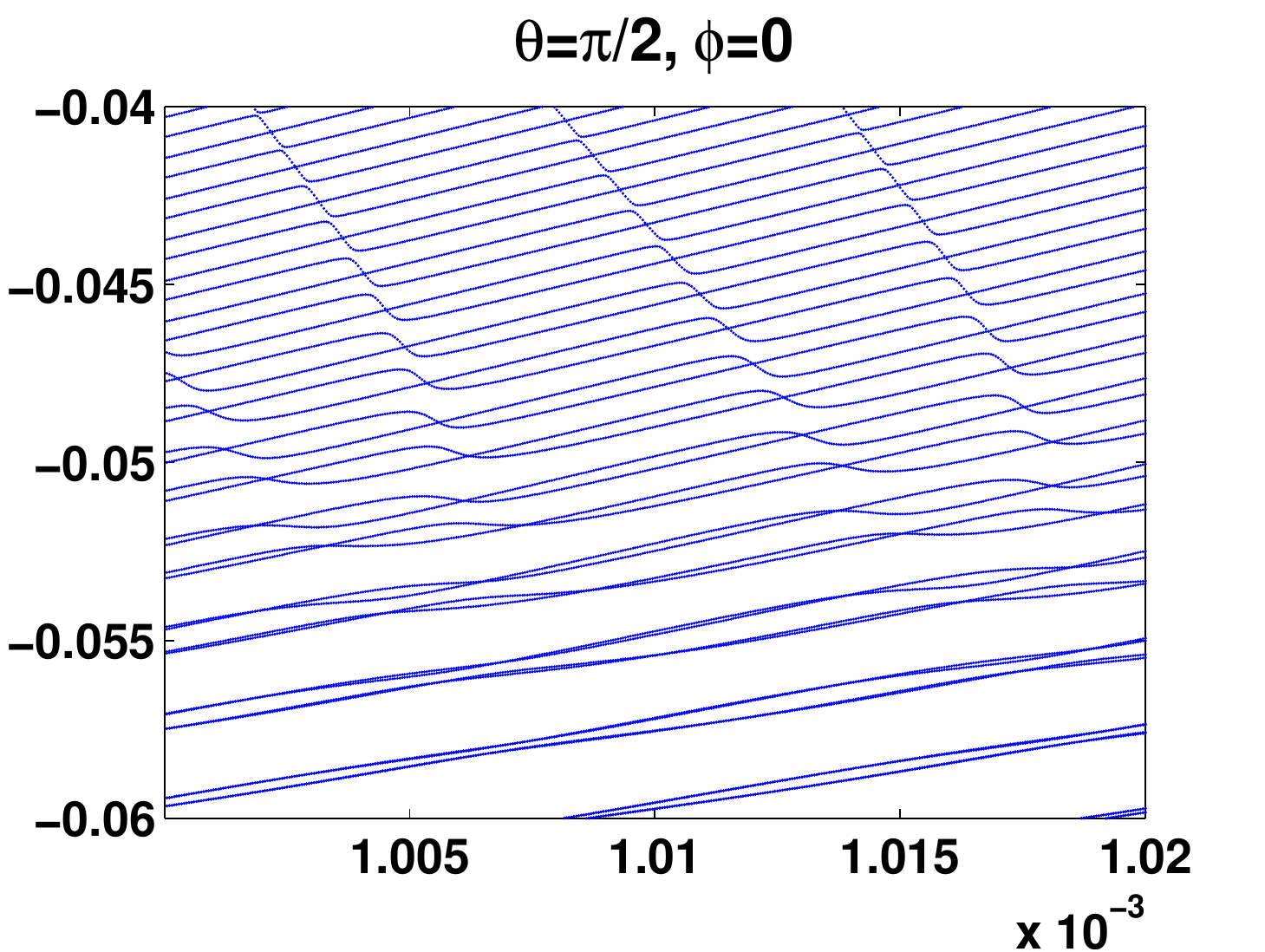}
\includegraphics[width=0.3\textwidth]{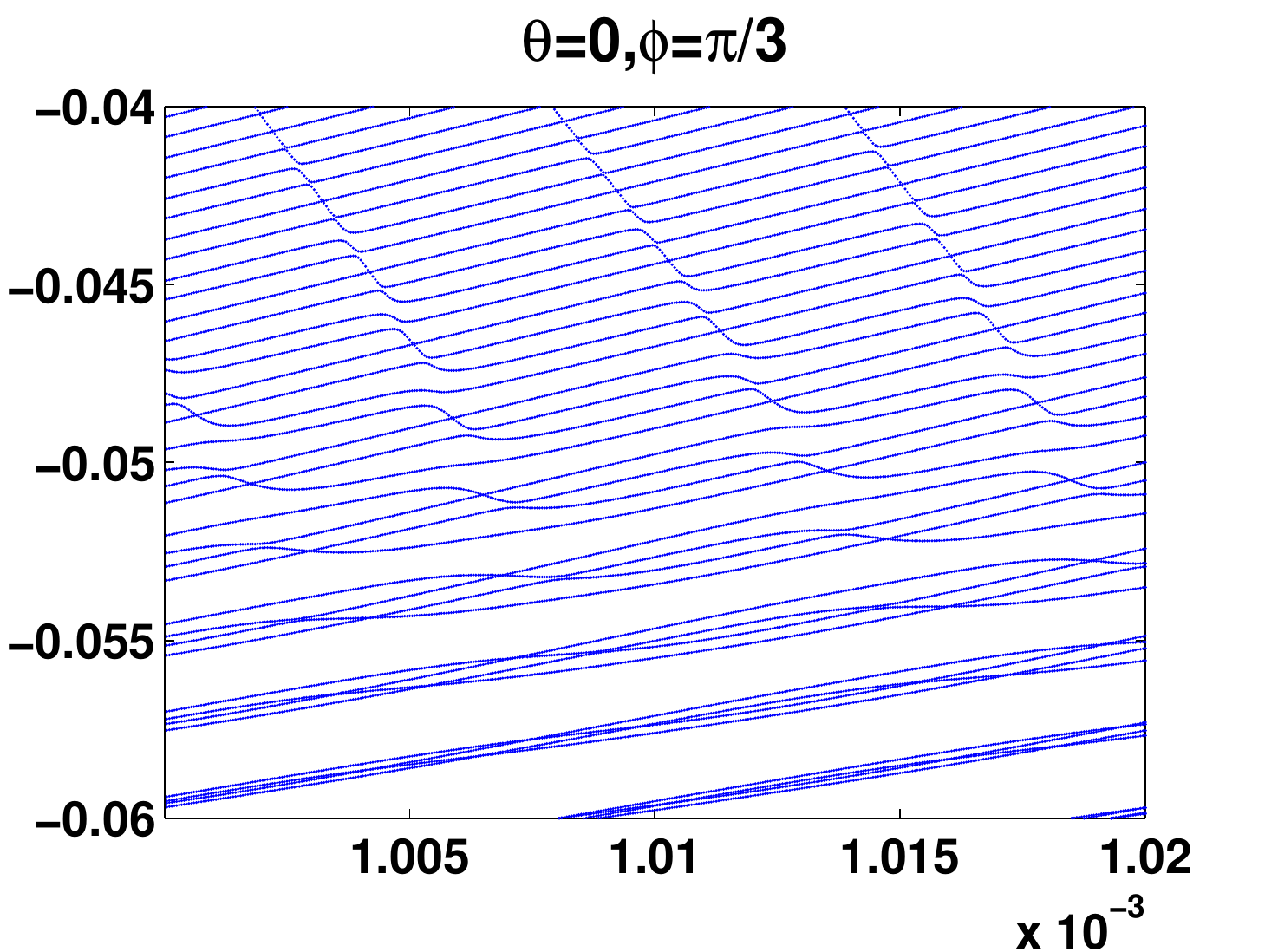}
\includegraphics[width=0.3\textwidth]{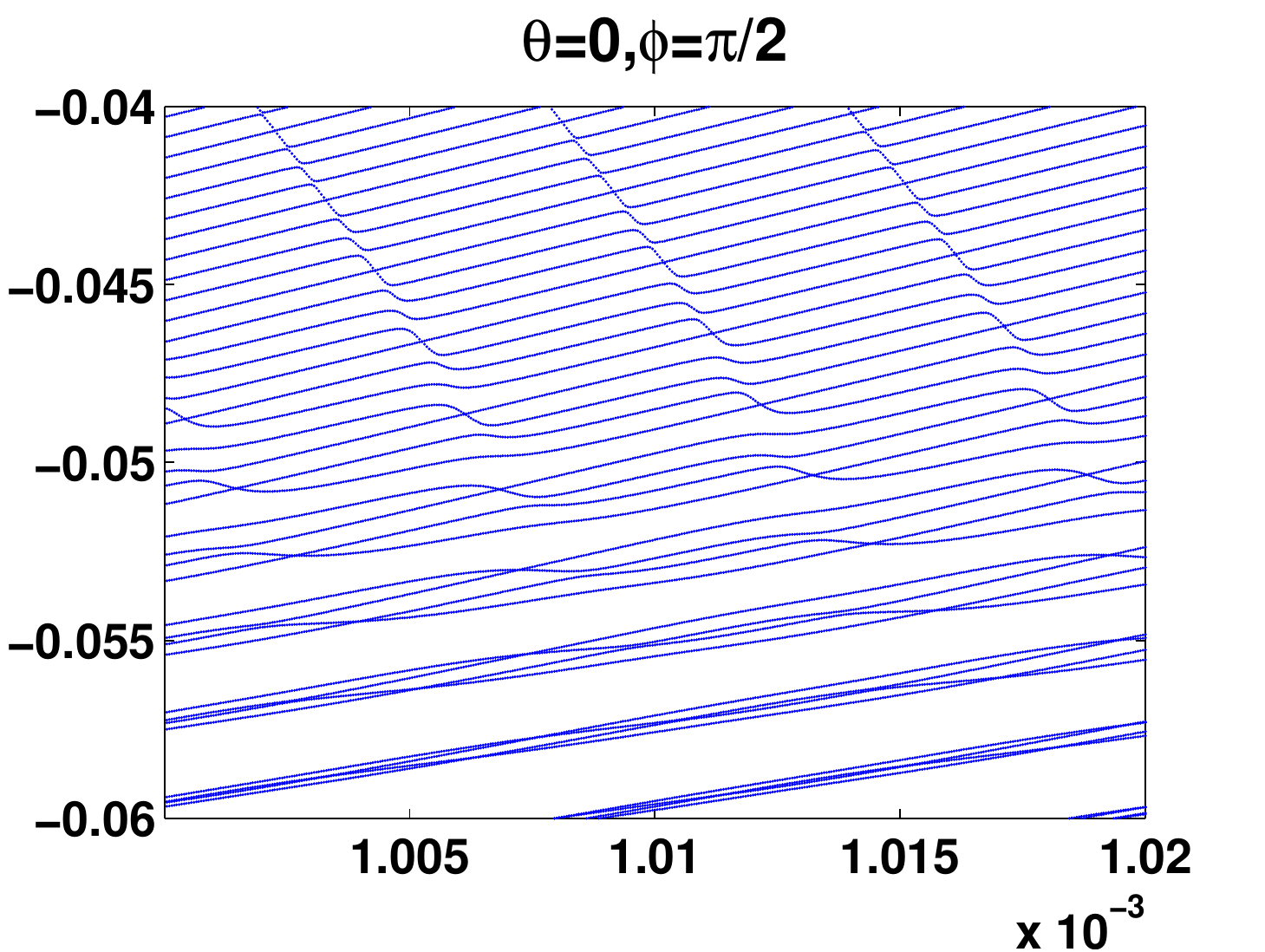}
\includegraphics[width=0.3\textwidth]{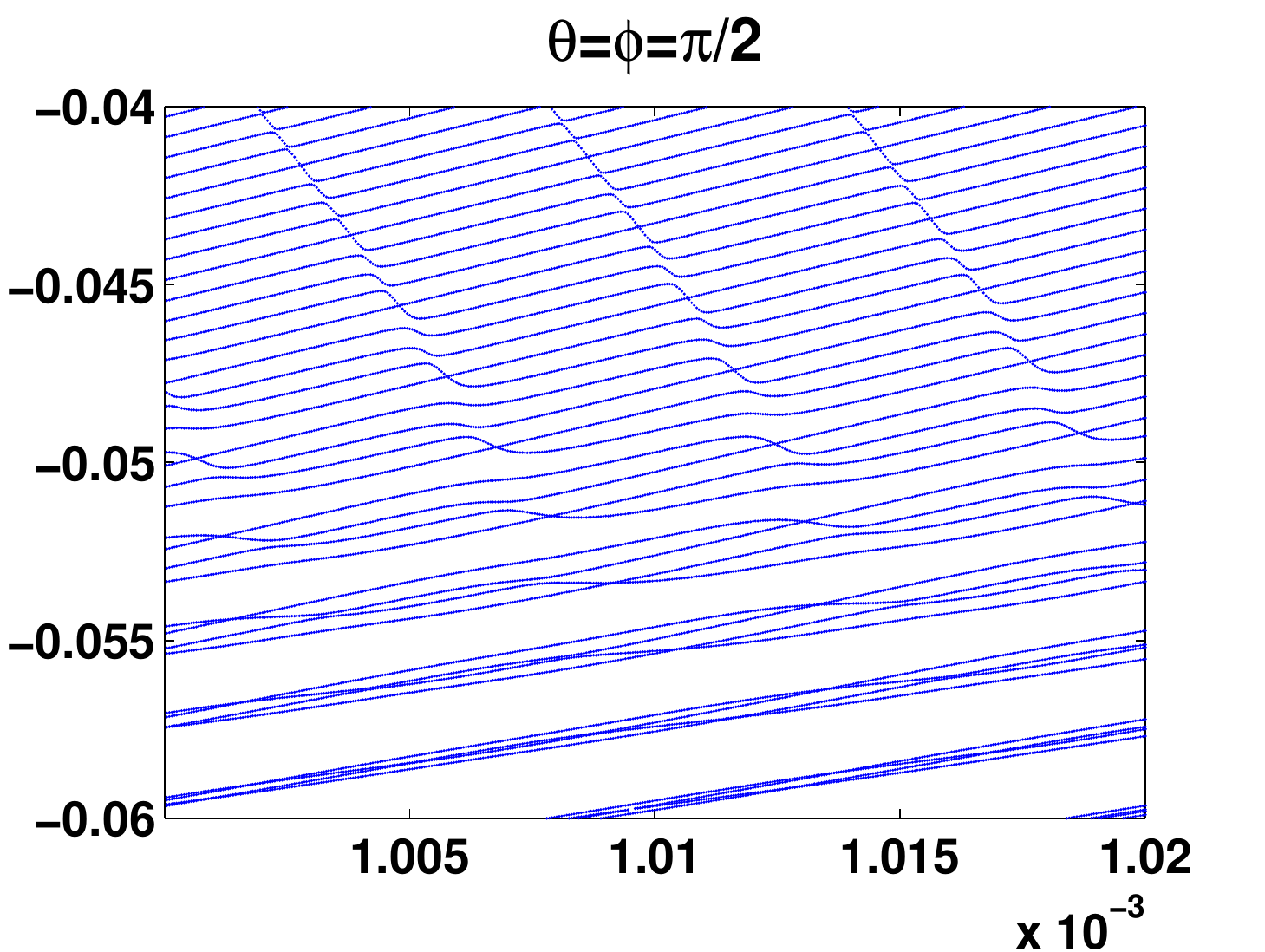}
\caption{Detailed views of the states near the saddle point for
various boundary conditions specified by the parameters $\theta$
and $\phi$. The spectrum away from saddle point at $E'=-0.05$ does
not change with these parameters.
}
\label{LL1}
\end{figure}


\begin{thebibliography}{plain}
\bibitem{footnote1} For example, had we chosen $\vec A=B(-2y/3,x/3)$,
one would need to retain three values of $\tilde k$ in the calculation.
The choice of circular gauge minimizes the number of $\tilde k$'s needed.
\end{thebibliography}
\end{document}